\documentclass[12pt]{iopart}

\expandafter\let\csname equation*\endcsname=\relax
\expandafter\let\csname endequation*\endcsname=\relax
\usepackage{multirow}
\usepackage{array}
\usepackage{float}

\usepackage{lineno}
\usepackage{amsmath,amssymb,amsthm}
\usepackage{iopams}
\usepackage{epsfig}
\usepackage{booktabs}
\usepackage{subfigure}
\usepackage{graphicx}
\usepackage{dcolumn}
\usepackage{stmaryrd}
\usepackage{mathrsfs}
\usepackage{pifont}
\usepackage{bm}
\usepackage{latexsym}
\usepackage{hyperref}
\usepackage{newfloat,algcompatible}
\usepackage[size=small]{caption}
\usepackage{etoolbox}
\usepackage{newfloat}
\usepackage{color}
\AtBeginEnvironment{algorithm}{\noindent\hrulefill\par\nobreak\vskip-5pt}
\DeclareFloatingEnvironment[fileext=loa,listname=List of Algorithms,name=Algorithm,placement=tbhp,]{algorithm}
\DeclareCaptionFormat{algorithms}{\vskip-15pt\hrulefill\par#1#2#3\vskip-6pt\hrulefill}
\pdfpageheight\paperheight
\pdfpagewidth\paperwidth
\UseRawInputEncoding
\makeatletter

\providecommand{\tabularnewline}{\\}
\floatstyle{ruled}
\newfloat{algorithm}{H}{loa}
\providecommand{\algorithmname}{Algorithm}
\floatname{algorithm}{\protect\algorithmname}
\makeatother
\usepackage{babel}
\providecommand{\tabularnewline}{\\}
\begin{document}
\title[]{Universal quantum state preparation via revised greedy algorithm}

\author{Run-Hong He, Hai-Da Liu, Sheng-Bin Wang, Jing Wu, Shen-Shuang Nie and Zhao-Ming Wang$^*$}

\address{College of Physics and Optoelectronic Engineering, Ocean University of China, Qingdao 266100, China}
\ead{mingmoon78@126.com}
\vspace{10pt}
\begin{indented}
\item[]April  2021
\end{indented}
\begin{abstract}
Preparation of quantum state lies at the heart of quantum information
processing. The greedy algorithm provides a potential method to effectively prepare quantum states. 
However, the standard greedy algorithm, in general, cannot take the global maxima and instead becomes stuck on a local maxima. Based on the standard greedy algorithm, in
this paper we propose a revised version to design dynamic pulses
to realize universal quantum state preparation, i.e., preparing any arbitrary state from another arbitrary one. 
As applications, we implement this scheme to the universal preparation of single- and two-qubit state
in the context of semiconductor quantum dots and superconducting circuits. Evaluation results show that our scheme outperforms
the alternative numerical optimizations with higher preparation quality
while possesses the comparable high efficiency.  Compared with the emerging
machine learning, it shows a better accessibility and does not require
any training. Moreover, the numerical results show that the pulse sequences generated by our scheme are robust against various errors and noises.
Our scheme opens a new avenue of optimization in few-level system and limited action space quantum control problems.
\end{abstract}

\vspace{2pc}

\section{Introduction \label{sec:introduction}}
	
Benefited from the fascinating abilities afforded by the quantum mechanics,
quantum computers of the future are supposed to tackle specific tasks
that are intractable or even prohibitive to solve with their classical
counterparts \cite{Nielsen_Quantum,quantum_computation_2020,shor_factoring,quantum_simulation1,quantum_simulation2,grover_searching,quantum_optimization,Wang_shengbin}.
Over the past few decades, a wide variety of physical modalities has been
proposed theoretically and demonstrated experimentally to construct
the prototypes of quantum computers, spanning from a single electron
to the topological system \cite{nmr,photonic_lattices1,photonic_lattices2,trapped_ions1,trapped_ions2,nv_center1,nv_center2,toric}.
Along with the mature of the hardware, the task
of designing control trajectory partially falls on the programming
side which bridges quantum science and traditional disciplines.

It has been proved that with only arbitrary single-qubit rotations
on the Bloch sphere plus an entangling two-qubit gate, arbitrary quantum
logic can be performed on a gate-based quantum
computer, or rather, they are universal \cite{Nielsen_Quantum}. 
Various scenarios have discussed how to decompose a quantum algorithm into an arrangement of these universal gates \cite{Nielsen_Quantum,cos_sin_decomposition,KAK_decomposition,krauss_cirac_decomposition}. 
Meanwhile, it has also been explored how to construct optimal
gate synthesis with the universal gate set for general logical operation
\cite{gate_synthesis}. While, for a given experimental platform,
there are always certain gates that are more efficient to implement
than others, or even the key ingredients of the latter. Thus, they
are often referred to as ``native gates'' of that platform. 
To enact quantum computation, it is required to
decompose a given quantum algorithm into a sequence of discrete native gates
according to the underlying hardware properties \cite{native_gate1,native_gate2}.

Experimentally, these native gates are performed by electromagnetic pulses with precise amplitudes and durations.
However, the difficulty of scheduling these pulses to implement a logical gate sharp increases with the
decrease of the degrees of freedom in general. A prominent example is the pluses designing for a singlet-triplet ($S$-$T_{0}$)
qubit in a semiconductor double quantum dot (DQD). Where the only
tunable parameter is the exchange coupling between two trapped electrons,
which associates with the rotation rate about the $z$-axis of the
Bloch sphere. The process of designing pulses analytically requires
to solve iteratively a set of nonlinear equations \cite{pulse_design_nonlinear,Wangxin_composite_pulses_qd,pulse_control_qd}.
Thus it is a overhead costly and time-consuming task in practice.
In order to improve efficiency, Ref.~\cite{pulse_design_supervided_learning}
discussed design pulses with supervised learning to get an approximation of that analytical
solution. Considering the challenge to realize pulses with continuous
intensity and duration in experiment, Ref.~\cite{zhangxiaoming_npj}
employed deep reinforcement learning \cite{RL_CNOT,RL_Universal,RL_Grover,RL_cartpole} to explore the preparation of
a specific state from another one with dynamic pulses whose intensity
and duration are both discrete, yet at the expense of universality.
Respecting the virtues of discrete control, Ref.~\cite{herunhong_2020_arxiv}
studied the preparation of a certain state from an arbitrary state.
In contrast, Ref.~\cite{ppo_state_preparation} promised to
prepare an arbitrary state from a specific state in a multi-level
nitrogen-vacancy center system. It is a meaningful point that by combining Refs.~\cite{herunhong_2020_arxiv} and \cite{ppo_state_preparation} the driving between any states can be implemented as suggested in Ref.~\cite{herunhong_2020_arxiv}. In addition, it is also a promising direction to train the network with both random initial state and target state to achieve the same objective directly.
Except for the nascent machine learning, there are also several sophisticated versatile optimization approaches
based on gradient can be utilized, such as stochastic gradient descent (SGD) \cite{SGD},
gradient ascent pulse engineering (GRAPE) \cite{GRAPE_1,GRAPE_2}
and chopped random-basis optimization (CRAB) \cite{CRAB_1,CRAB_2}.
They have been successfully applied to a wide array of optimization problems.
While, suffering from the sensitivity to the initial control trajectory setting,
in general, they can find only local maxima, instead of global maxima,
and quit the iterative process with an inadequate fidelity.   

In this paper, based on the standard greedy (SG) algorithm \cite{greedy_algorithm1,greedy_algorithm2}, a common technology for optimization, 
we provide an improved version, i.e. revised greedy (RG) algorithm
to drive an arbitrary state to another arbitrary state, or say, universal
state preparation with discrete control. On the one hand, differing
from the algorithms based on the machine learning, which suffers from
the long hours of training and the resulting huge computational overhead,
our scheme needs no training at all, which ensures a high accessibility. On the other hand, contrasting to the traditional
optimization methods, our scheme overcomes the local optimality and
achieves a higher preparing quality. In addition, compared with
them, the average runtime of identifying the appropriate pulses with
our scheme is comparable to the GRAPE, which
is known for its high efficiency. We apply our scheme exemplarily
to the universal single- and two-qubit state preparation in the context
of $S$-$T_{0}$ spin qubits in semiconductor DQDs and superconducting
quantum circuits Xmon qubits \cite{x-mon_paper,x-mon_1}. Our
method is general enough to be extended to varieties of few-level system and limited action space quantum control problems.

The remainder of this paper is organized as follows. In Sec.~\ref{sec:method},
we describe in detail the method used in this work. Then
we present the results in the Sec.~\ref{sec:results}, and
conclude in Sec.~\ref{sec:conclusion}. The models considered in this paper are collected in Sec.~\ref{sec:models}.
	
\section{Method \label{sec:method}}
Dynamic programming, an important member of the optimization theory,
divides a complex problem into multiple simple sub-problems, and then
solves each sub-problem individually. It is expert in solving the Markov
decision process, in which the resulting state $S'$ is determined
uniquely by the current state $S$ and the action $a$ taken by the
agent while has no connection with the history of the system \cite{reinforcement_learning_sutton}.
The SG algorithm, which is build upon the dynamic programming, approximates the
global result by collecting the optimal solutions of each sub-problem
\cite{greedy_algorithm1,greedy_algorithm2}.
The SG algorithm as well as its variants is the most commonly used strategy to determine which action to explore in a given state in optimizations and acts as a crucial ingredient in many successful quantum control schemes, such as the designing of high-fidelity quantum gates \cite{Toffoli_gate,Control_Z_gate} and the scheduling of quantum gates to implement quantum algorithm \cite{Scheduling_gates}.
However, because of the overemphasis on local optimality, in general, the SG algorithm cannot produce the global maxima. 
In order to overcome the local optimality of the SG algorithm, we propose
here an revised version, i.e. RG algorithm to design the control pulses.

\begin{figure}[htb] 
\centering
\includegraphics[width=8cm]{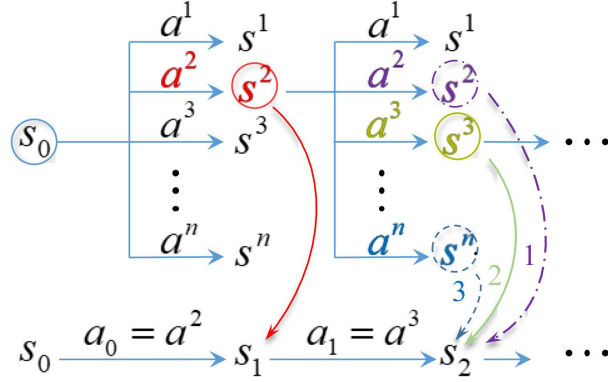}
\caption{Schematic of the RG algorithm for control trajectory designing. The details of this algorithm are described in the main text of Sec.~\ref{sec:method} and in the pseudocode
Algor.~\ref{algorithm1}.}
\label{Fig1}
\end{figure}

Our target is to design pulses to drive an arbitrary given quantum
state to another arbitrary state.
To illustrate the process of designing pulses clearly, we consider exemplarily
the driving from the initial state $|S_{0}\rangle=|0\rangle$ to the
target state $|S_{tar}\rangle=|1\rangle$ with single-parameter dynamic pulses. The quality of the state preparation is evaluated by
the fidelity, which is defined as $F\equiv|\langle S_{tar}|S_{n}\rangle|^{2}$,
where $|S_{n}\rangle$ (also noted as $S_{n}$ for simplicity) refers
to the evolution state at time step $n$. The schematic of this processing
is patterned in Fig.~\ref{Fig1}. 
To reduce the computational overhead, the control function is discretized as a piecewise constant (PWC) pulse-sequence  \cite{GRAPE_2}. The maximum evolution time $T$ is divided uniformly into $N$ slices with pulse duration $\mathrm{d}t=T/N$. 
All of the allowed actions adopted in this work accommodate the fundamental constraints and experimental realities, such as pulse height, duration, etc \cite{coherent_manipulation_qd,quantum_dot_singlet-triplet_2_qubit,superconducting_guide}.  
This control function can be readily implemented with suitable electrodes voltages generated by an arbitrary waveform generator in the platform of semiconductor QDs \cite{quantum_dot_singlet-triplet_electrically-control2}. While, for the control of superconducting circuits, these pulses with discrete intensity and duration can be translated into continuous microwaves generated by a typical IQ modulation setup \cite{superconducting_guide}.
Presume there are four allowed
actions $\{a^{1},a^{2},a^{3},a^{4}\}$ corresponding to four
allowed pulse strengths respectively.

The scheme goes as follows: initially, the fidelity of the state
$S_{0}$ is $F_{0}=0$. Then, calculate each fidelity $F^{i}$
caused by the corresponding action $a^{i}$ after $\mathrm{d}t$ evolution.
We assume these resulting fidelities are 0.1, 0.3, 0.2, 0.15 respectively,
and all of them are bigger than $F_{0}$. Then we choose the maximum
of $F^{i}$ as the fidelity at this time step, i.e. $F_{1}\leftarrow F^{2}=0.3$;
the corresponding action as the ``selected action'', $a_{0}\leftarrow a^{2}$;
and the corresponding evolution state as the next state, $S_{1}\leftarrow S^{2}$.
This step dovetails neatly with the SG algorithm: employ directly
the ``best action''. After, based on the state $S_{1}$, perform the
allowed actions $a^{i}$ separately again and take the corresponding
fidelities $F^{i}$. Assume the resulting fidelities are 0.26, 0.3,
0.28 and 0.25 respectively, and obviously there's no new fidelity bigger
than $F_{1}$. To overcome the local optimality in this case, we employ
3 strategies to decide which action should be selected: \textbf{strategy
1}, choose the ``best action'' according to the greedy algorithm;
\textbf{strategy 2}, the ``next-best action''; and \textbf{strategy
3}, the ``worst action''. And repeat the above operations, until the
final time step $N$ or the fidelity exceeds a certain satisfactory
threshold. Note that in an episode, from the first step to the
last, only one strategy is adopted to ensure the stability of the
algorithm. Finally, take the action-sequence as the solution corresponding
to the strategy in which we obtained the maximum fidelity.
For current computer, with more than enough CPU cores available, these strategies 
can be readily executed on different processings in parallel, making it an extremely effective algorithm. 
The pseudocode of this RG algorithm is given in Algor.~\ref{algorithm1}.

The core of this scheme lies at the introduction of deliberate perturbation, 
represented as the selection of ``non-best'' action, when the state are stuck on local maxima, 
which is regarded empirically as an efficient way to  achieve a better performance \cite{GRAPE_2}. 
This setting is necessary for cases like that: with the north pole of Bloch sphere targeted, 
the ``best'' action may always orient the state located in the equator along the equator 
and cannot reach a better place forever. 
In addition, unlike algorithms based on machine learning, there's no training at all.
Alternatively, the RG explores the suited action as well as the next state by trials and errors online, 
converting a complicated model-free environment into a simple model-based one which the algorithm 
fully understands \cite{reinforcement_learning_sutton}. 
So that it does not require learning of the environment and actions any more. 

Our scheme achieves a better performance than the comparable approaches in few-level system and limited action space quantum control problems, such as the single- and two-qubit state preparation as we will demonstrate in the following section. Nonetheless, this advantage may diminish as the number of qubits and the action space blow up, partially due to the huge increase in them will limit this method's efficiency. (For optimization problems with continuous action space, the machine learning certainly has more natural advantages.)   In addition, in many-body preparation, such as the quantum state transfer, there may be a lack of a well metric to determine which state is ``better" than others in the intermediate process. One possible route for improvement is to use this method in concert with other algorithms, which we leave as future works. However, even if the many-body state preparation is a very important topic from the application point of view, as we mentioned in the Introduction, with only arbitrary single- and two-qubit gates, any quantum logic can be implemented on a circuit-model quantum computer.

\begin{algorithm}
\caption{Pseudocode of the RG algorithm for control trajectory designing.}

\label{algorithm1}
\begin{algorithmic}
\STATE Initialize the time step $step=0$ and state $s_{step}$ according
to the testing point.
\STATE Calculate the fidelity $F_{step}$ and make $F_{\mathrm{max}}=F_{0}=F_{step}$.
\WHILE{True} 

\STATE Perform the allowed actions $a^{i}$ separately and record the corresponding
fidelity $F^{i}$.
\STATE \textbf{If} $\mathrm{max}(F^{i})>F_{step}$, let $F_{step+1}\leftarrow\mathrm{max}(F^{i})$,
specify the corresponding action $a_{step}=\mathrm{argmax}_{a^{i}}F^{i}$
as the ``\textsl{best action}'' in this time step and make $F_{\mathrm{max}}\leftarrow F_{step+1}$.
\STATE \textbf{Else}, assign the \textsl{``best action}'' (strategy 1)
or \textquotedblleft \textsl{next-best}\textit{ action}\textquotedblright{}
(strategy 2) or ``\textsl{worst action}'' (strategy 3) as the \textit{``selected
action}'' $a_{step}$ and record the corresponding fidelity as $F_{step+1}$.
\STATE Next state $s_{step+1}$ are the state caused by performing the \textit{``selected
action}''.
\STATE Let $s_{step}\leftarrow s_{step+1}$ and $step=step+1$.
\STATE \textbf{Break} if $step\geqslant step_{\mathrm{max}}$ or $F>0.999$.
\ENDWHILE
\STATE Output $F_{\mathrm{max}}$ as the maximum fidelity and the action sequence
composed of the ``\textit{selected actions}'' from $step=0$ to
the $step_{end}$ in which we take the $F_{\text{\ensuremath{\mathrm{max}}}}$.
\end{algorithmic}
\end{algorithm}

\section{Results \label{sec:results}}

In the preceding section, we have presented the method used in this work. Now, we consider four cases of state preparation: single- and two-qubit state preparation in the context
of semiconductor DQDs or superconducting circuits. The details of the models are introduced in Sec.~\ref{sec:models}.

Any single-qubit state can be graphical represented by a point on
the Bloch sphere
\begin{equation}
|\Psi(\theta,\varphi)\rangle=\mathrm{cos}(\frac{\theta}{2})|0\rangle+\mathrm{e}^{i\varphi}\mathrm{sin}(\frac{\theta}{2})|1\rangle,
\end{equation}
where the polar angle $\theta\in[0,\pi]$ and the azimuthal angle
$\varphi\in[0,2\pi)$ \cite{Nielsen_Quantum}. To verify the universality,
for the single-qubit state preparation, we sample 128 testing points
on the Bloch sphere, which is distributed uniformly at the angles $\theta$
and $\varphi$.

For the two-qubit state preparation, we take a data set comprising
6912 points which are defined as $\{[a_{1},a_{2},a_{3},a_{4}]^{T}\}$,
where $a_{j}=\mathrm{e}^{i\phi}c_{j}$ refers to the probability amplitude
of the $j$th basis state, with $\phi\in\{0,\pi/2,\pi,2\pi/3\}$;
and these $c_{j}$s together indicate the coordinates of points scattered
on a 4-dimensional unit hypersphere
\begin{equation}
\begin{cases}
\begin{array}{ll}
c_{1}= & \mathrm{cos}\theta_{1},\\
c_{2}= & \mathrm{sin}\theta_{1}\mathrm{cos}\theta_{2},\\
c_{3}= & \mathrm{sin\theta_{1}\mathrm{sin}\theta_{2}\mathrm{cos}\theta_{3},}\\
c_{4}= & \mathrm{sin}\theta_{1}\mathrm{sin}\theta_{2}\mathrm{sin}\theta_{3},
\end{array}\end{cases}\label{eq:7}
\end{equation}
with $\theta_{i}\in\{\pi/8,\pi/4,3\pi/8\}$ \cite{herunhong_2020_arxiv}.
While, to reduce the overhead, we select randomly 512 samples from
that data set to form the testing set in this case.

Each point in the testing set will be prepared as target state from
all other points in turn. And then the average fidelity $\overline{F}$
of each target state preparation can be taken. The universality is
evaluated by the mean of these average fidelities $\langle\overline{F}\rangle$
over all target states: for the cases of single-qubit state preparation,
there are $128\times(128-1)=16,256$ preparation tasks. For
the two-qubit cases, the total number of tasks is $261,632$.

\subsection{Universal single-qubit state preparation with revised greedy algorithm}

Arbitrary manipulations of a single-qubit state can be achieved by successive
rotations on the Bloch sphere, which are completed by a sequence of control
pulses \cite{pulse_control_qd}. The only tunable parameter of single-qubit
in $S$-$T_{0}$ DQD is the coupling strength $J(t)$, which is bounded
physically to be non-negative and finite. Here, we take four discrete
allowed actions, i.e. $J\in\{0,1,2,3\}$, to drive single-qubit states.

In Xmon superconducting circuits system, the drives on $x$- and
$y$-directions are limited to be finite, while the drive on $z$-direction
is further restricted to be non-negative. We choose 11 discrete allowed
actions $A_{x(y)}\in\{-2,-1,0,1,2\}$, $A_{z}\in\{0,1,2\}$ and within
a time step we just take the action $A_{x}$, $A_{y}$ or $A_{z}$ alone.

\begin{figure}[htb]
\centering
\subfigure[ ]{\includegraphics[width=7.5cm]{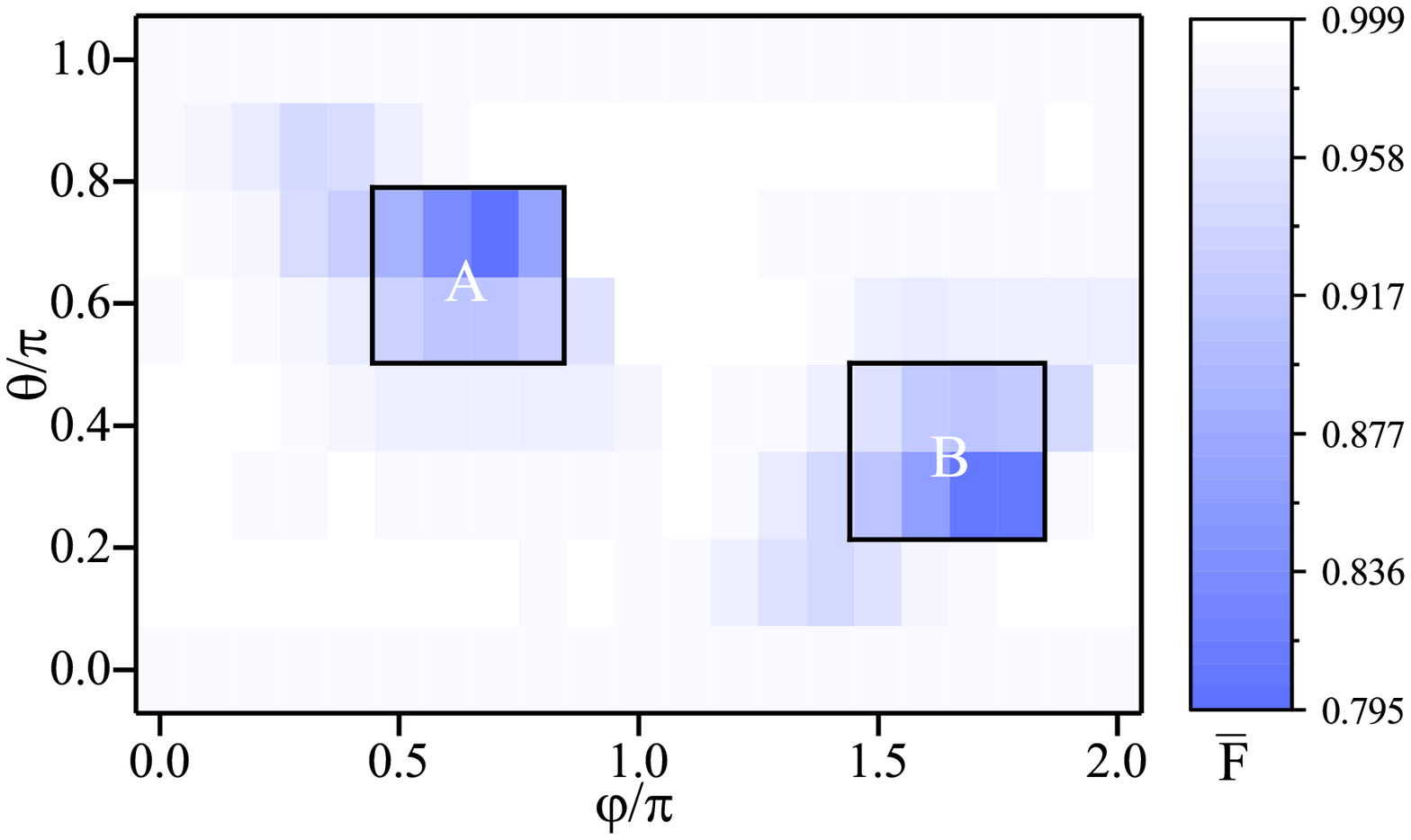}}
\subfigure[ ]{\includegraphics[width=7.5cm]{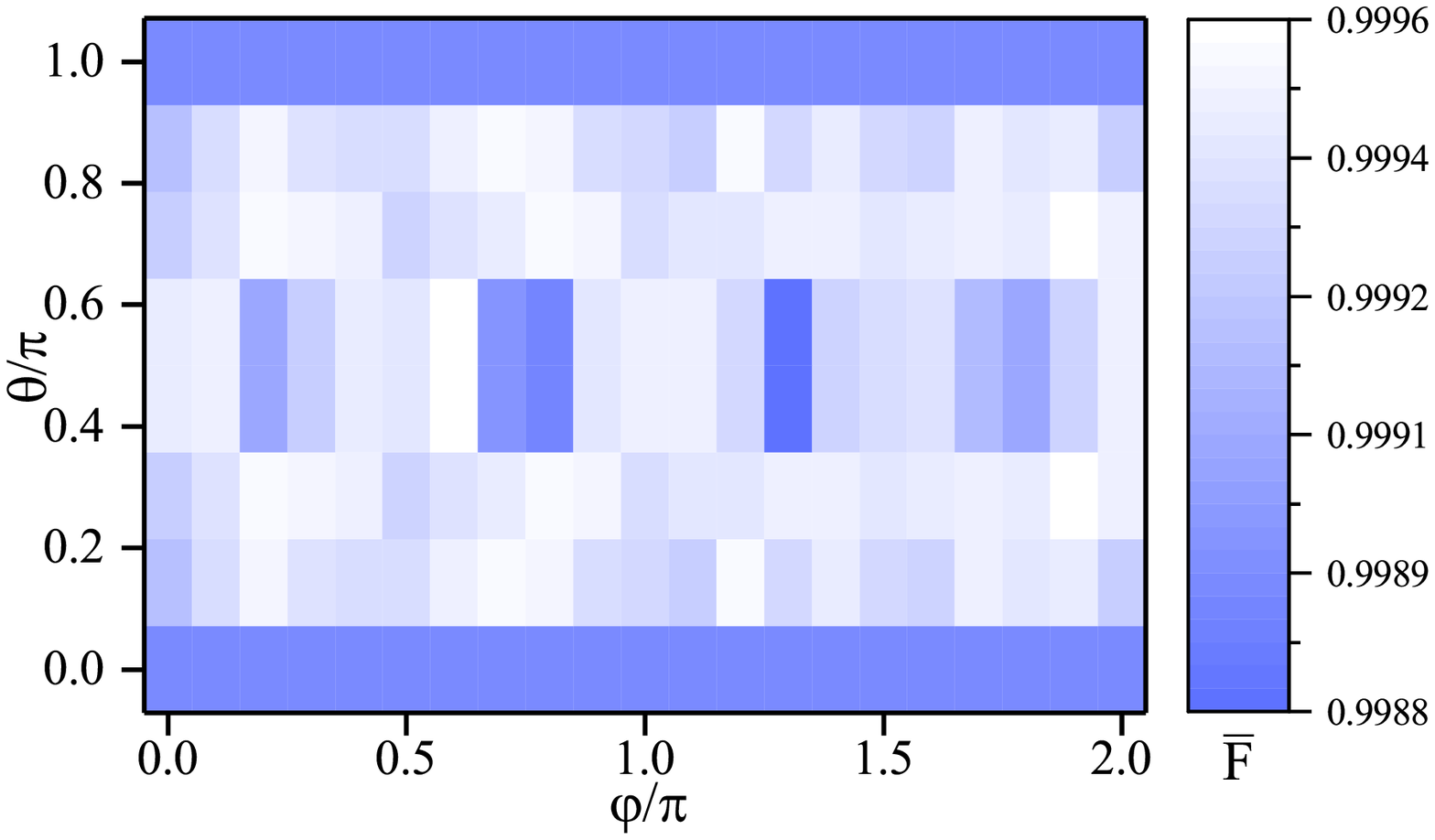}}		
\caption{Average fidelities $\overline{F}$ distribution of preparing arbitrary
single-qubit target state $\Psi(\theta,\varphi)$ using RG algorithm
over 127 sampled tasks. (a) The mean of all average fidelities $\langle\overline{F}\rangle=0.973$
in DQDs system with $T=2\pi$ and $\mathrm{d}t=\pi/5$. The areas A and B are clustered bad points whose mean of average fidelities $\langle\overline{F}\rangle$ are 0.902 and 0.886 respectively. (b) The mean
of all average fidelities $\langle\overline{F}\rangle=0.999$ in superconducting
circuits system with $T=\pi$ and $\mathrm{d}t=\pi/5$. \label{fig:2}}
\end{figure}

\begin{figure}[htb]
\centering
\subfigure[ ]{\includegraphics[width=7.5cm]{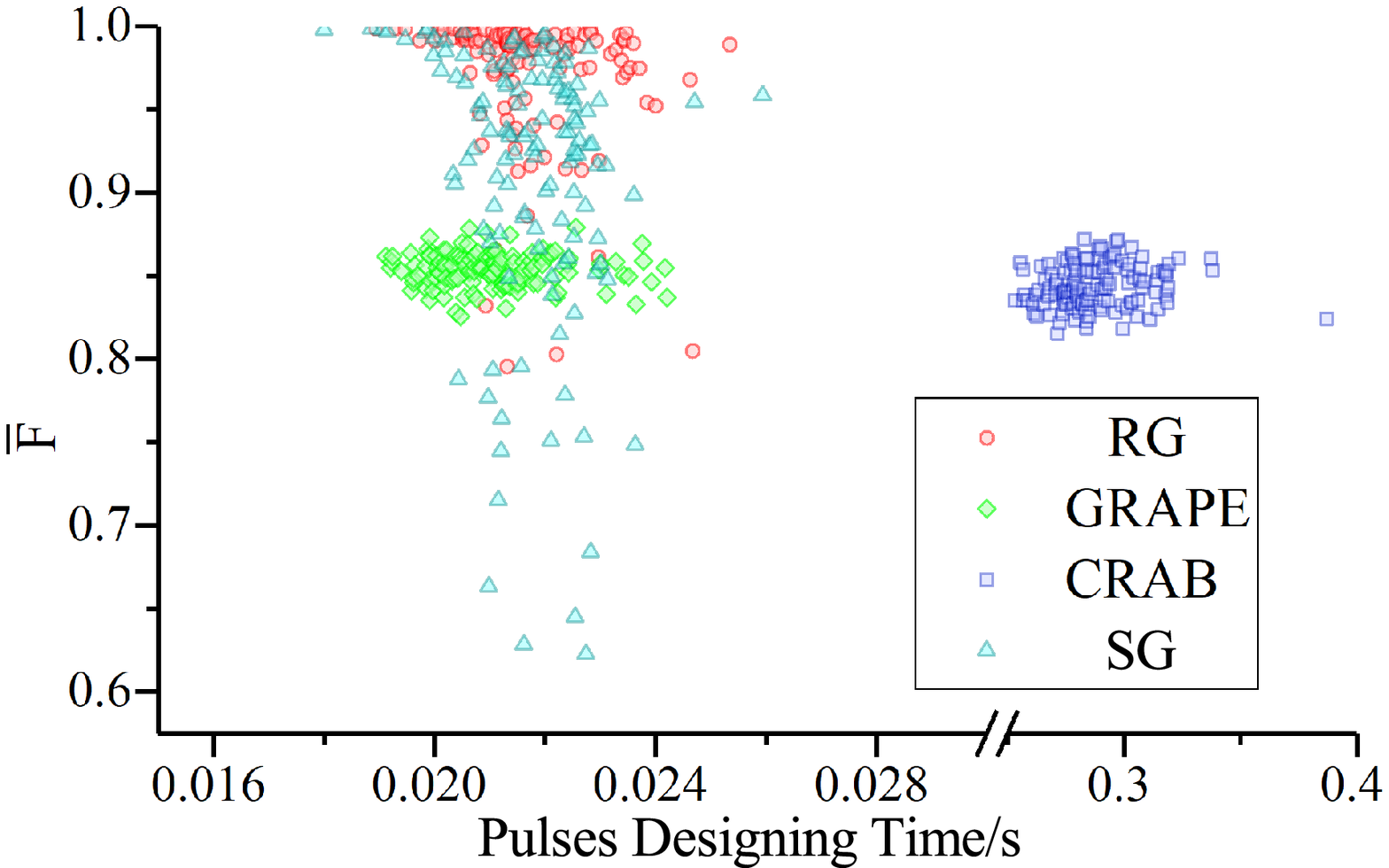}} 
\subfigure[ ]{\includegraphics[width=7.5cm]{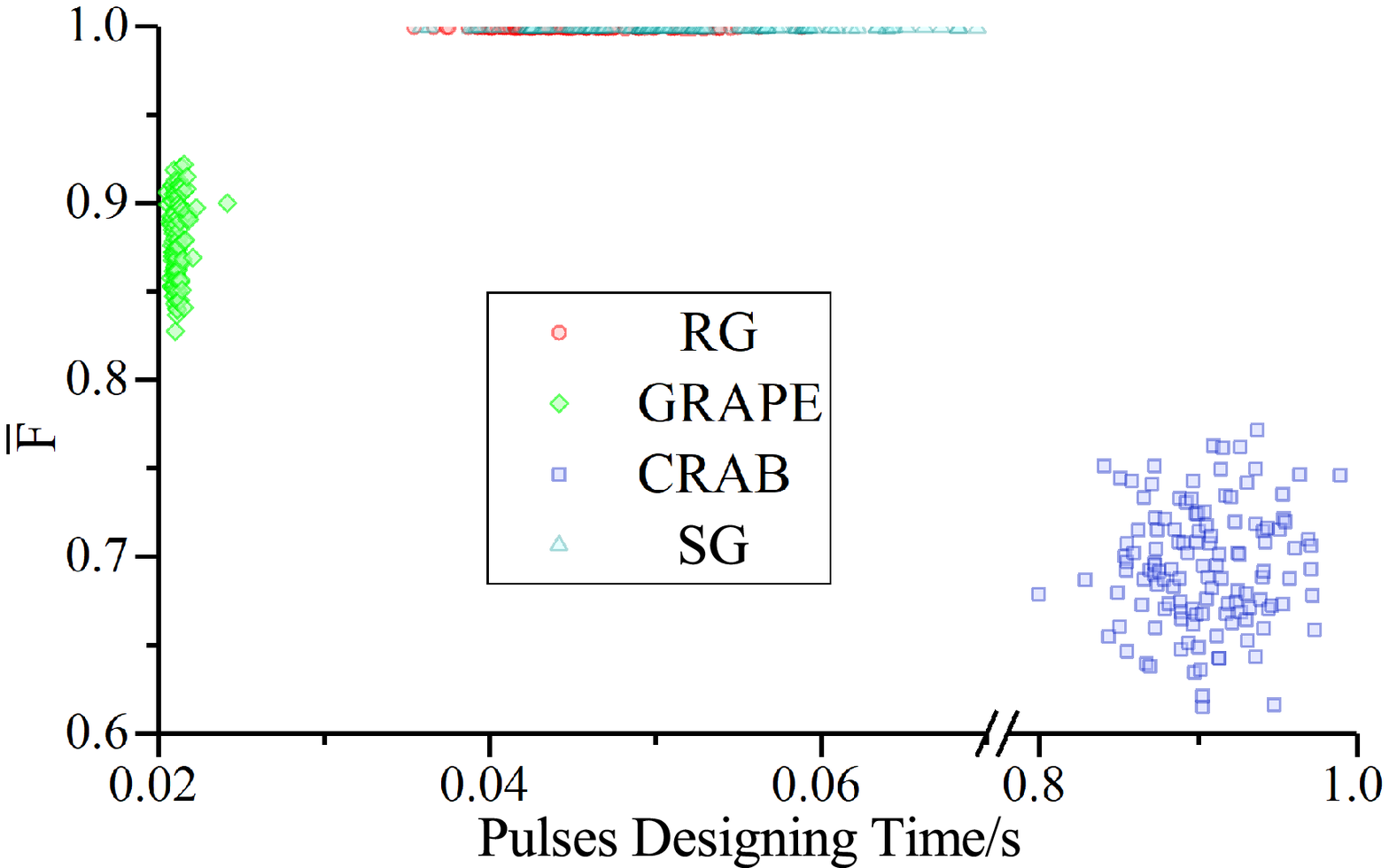}}
\caption{Average fidelities $\overline{F}$ versus designing time distributions of
preparing arbitrary single-qubit target state over 127 sampled tasks
with different optimization algorithms. The control parameters are
identical to that adopted in Fig.~\ref{fig:2}. (a) State preparation in $S$-$T_{0}$ DQD, where the mean of all average
fidelities $\langle\overline{F}\rangle=0.97273$, $0.85372$, $0.84407$, $0.91143$ and the
mean of all average designing time $\langle\overline{t}\rangle=0.0217$, $0.0211$, $0.2877$, $0.0217$
with RG, GRAPE, CRAB and SG. (b) States
preparation in Xmon superconducting circuits, where the mean of
all average fidelities $\langle\overline{F}\rangle=0.99944$, $0.87807$, $0.69425$, $0.99937$
and the mean of all average designing time $\langle\overline{t}\rangle=0.0449$, $0.0214$, $0.9050$, $0.0520$
with RG, GRAPE, CRAB and SG.\label{fig:3}}
\end{figure}

The results of state preparation under different
control parameters in two models are captured and shown in Table.~\ref{tab1}.
For visualization, Fig.~\ref{fig:2} (a) and (b) plot the average
preparation fidelity $\overline{F}$ of each single-qubit target state
parameterized by angles $\theta$ and $\varphi$ in the context of
semiconductor DQD and superconducting circuits respectively. The data
correspond to the second and tenth rows of the Table.~\ref{tab1},
respectively. In Fig.~\ref{fig:2} (a), we can see that, although
there's only one degree of freedom, the fidelities of state preparation
are high in most areas. Whereas there also some ``bad points''
exist and cluster together that cannot reach the target states well,
e.g., the areas A and B, where the corresponding $\langle\overline{F}\rangle=0.902$
and $0.886,$ respectively. We ascribe this partially to the inappropriate
parameter choice and believe it can be improved by further specifically
tailoring parameters in these areas such as extended evolution time,
altered action duration or more allowed actions. For example, leaving
the other parameters intact, when $T=8\pi$, $\mathrm{d}t=\pi/3$,
$\langle\overline{F}\rangle=0.992$ in both the A and B areas. In
contrast, benefited from the additional degrees of freedom in $x$-
and $y$-axes, the performance of state preparation in superconducting
circuits is much better than in DQD, as shown in Fig.~\ref{fig:2}
(b), whose $\langle\overline{F}\rangle=0.999$ and the minimum $\overline{F}$ still exceeds $0.998$.

In addition, for comparing the performance of our scheme with the
traditional optimization approaches, we plot the fidelities versus the corresponding
runtime of designing pulses of our scheme plus the GRAPE, CRAB and SG
for state preparation in DQDs and superconducting circuits in Fig.~\ref{fig:3}
(a) and (b) respectively, of which the control parameters are identical
to that adopted in Fig.~\ref{fig:2}. To ensure a fair comparison,
for GRAPE and CRAB, we discretize their continuous control strengths to the nearest allowed
actions at the end of the execution to get the final solution. It can
be seen that, our RG algorithm outperforms the GRAPE, CRAB, and SG with
higher quality of state preparation in both $S$-$T_{0}$ DQD and
Xmon superconducting single-qubit models. While, the runtime of
designing proper control trajectory is in the same order of magnitude
as the sophisticated GRAPE, which is known for high efficiency. 

It is worth stating that, the reasons are different for the diversity of designing time in different algorithms: the runtime of GRAPE and CRAB are mainly brought about by the number of iterations; while the runtime of SG and RG are mainly caused by the minimum time steps to finish an episode. That is, the time steps in GRAPE and CRAB are fixed - $N=T/\mathrm{d}t$; yet the time steps in SG and RG are alterable, which may be smaller than $N$ only if the requirement to terminate the episode early is met. (See the details of the RG described in Sec.~\ref{sec:method}.)
The RG favors a shorter path on the Bloch sphere between the initial and target states compared to other algorithms. Whereas, geometrically, this discrete control does not suffice to hit the quantum speed limit. We intend to make an another improved version of the SG utilizing pulses with continuous strength and duration to explore the speed limit of the similar quantum control problems in the future works.

\subsection{Universal two-qubit state preparation with revised greedy algorithm}

The control space for two-qubit state preparation in semiconductor
DQDs is parameterized by the allowed pulse strengths
in each qubit, i.e. $\{(J_{1},J_{2})|J_{1},J_{2}\in\{1,2,3,4,5\}\}$.
Thus, there are $5\times5=25$ allowed actions. As for two-qubit states
preparation in superconducting circuits, the allowed actions on each
qubit are same as these taken in the single-qubit case. The total
number of allowed actions is $11\times11=121$.

\begin{figure}[htb]
\centering
\subfigure[ ]{\includegraphics[width=7.5cm]{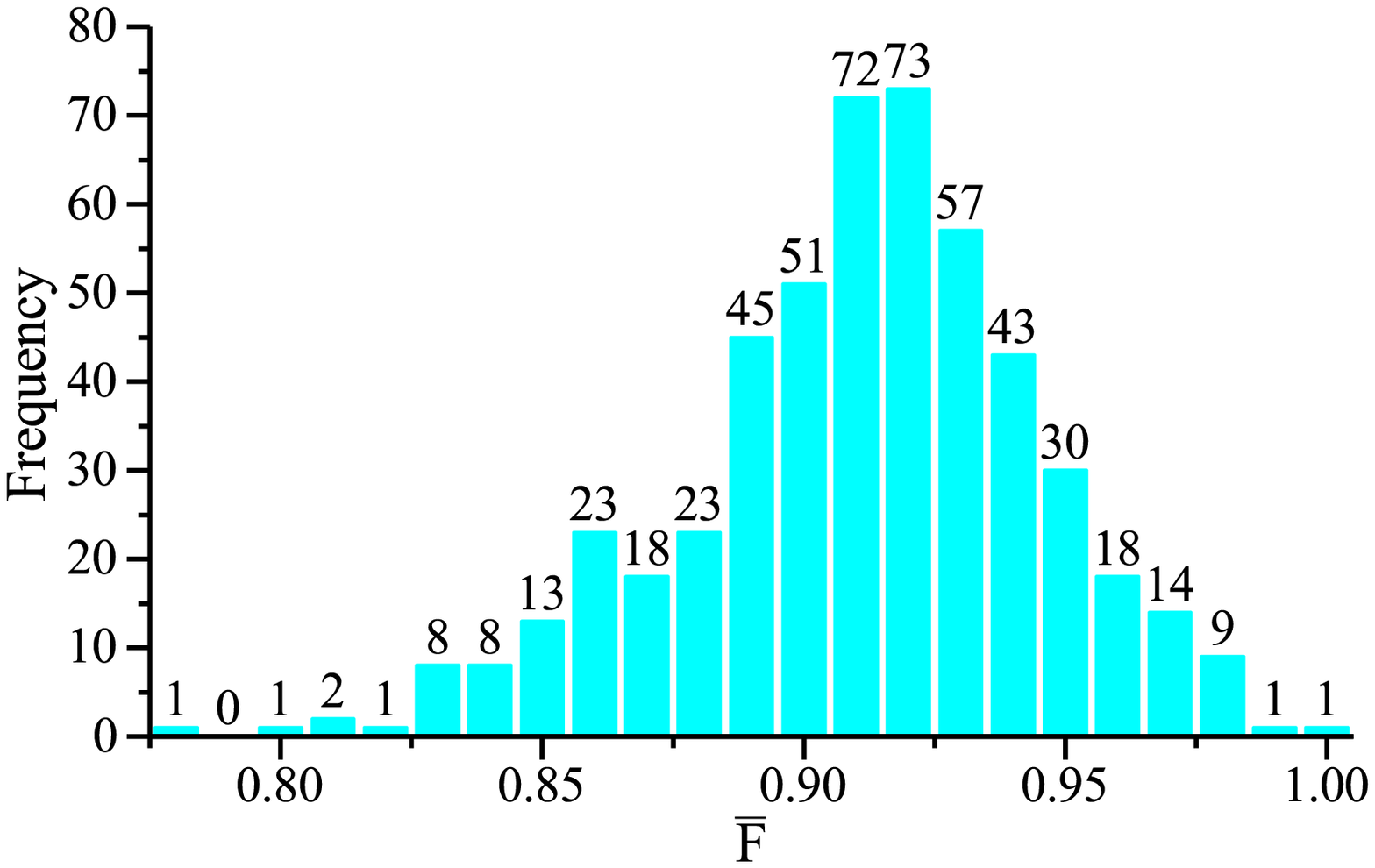}}
\subfigure[ ]{\includegraphics[width=7.5cm]{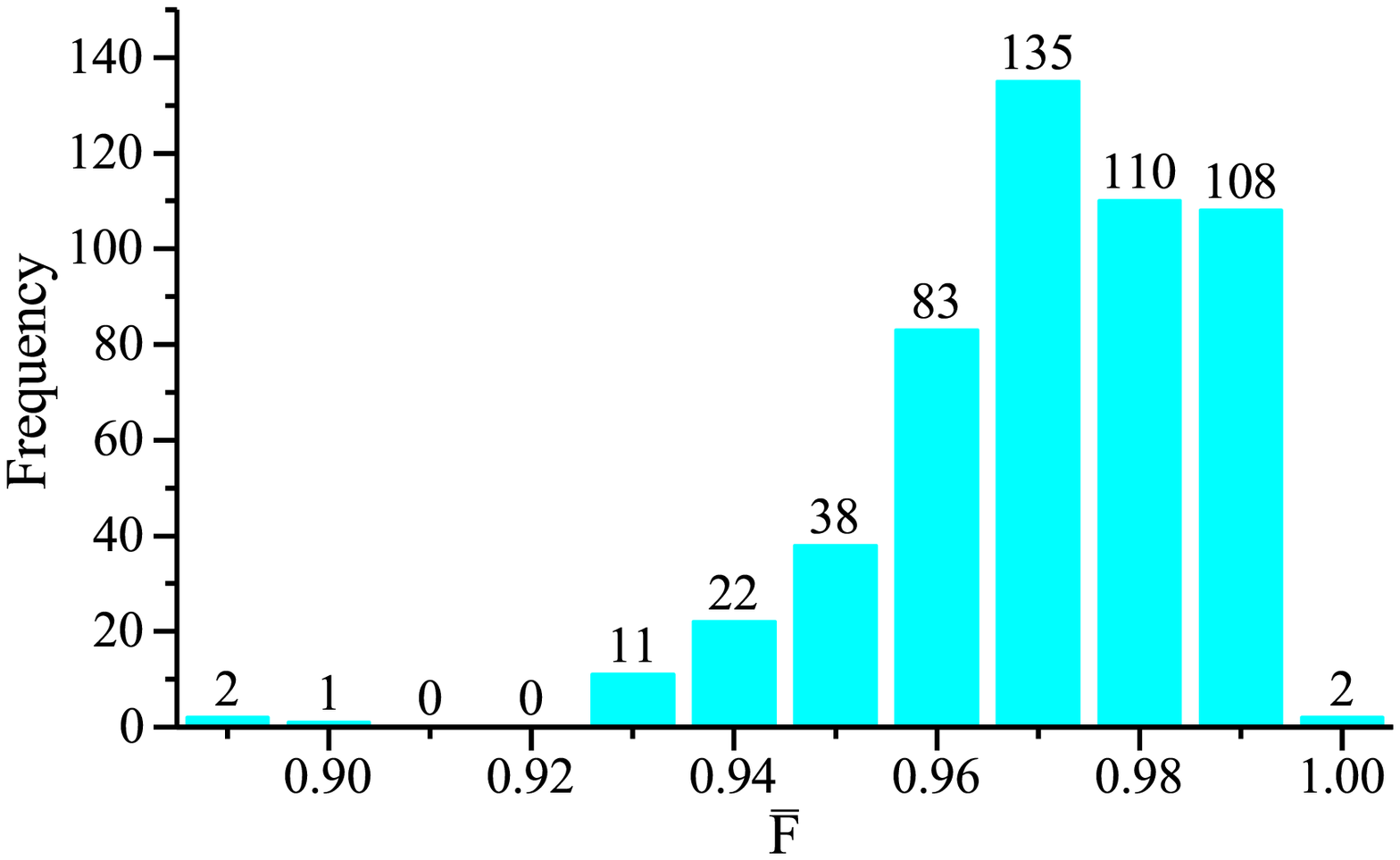}}
\caption{The frequency distributions of average fidelities $\overline{F}$ for
two-qubit state preparation using RG algorithm over
511 sampled tasks. (a) The mean of all average fidelities $\langle\overline{F}\rangle=0.911$
in DQDs system with $T=10\pi$ and $\mathrm{d}t=\pi/2$. (b) The mean
of all average fidelities $\langle\overline{F}\rangle=0.971$ in superconducting
circuits system with $T=10\pi$ and $\mathrm{d}t=\pi/4$. \label{fig:4}}
\end{figure}

The detailed control parameters and the corresponding results in two
models are listed in Table.~\ref{tab1}. Meanwhile, Fig.~\ref{fig:4}
(a) and (b) plot the frequency distributions of average preparation
fidelity $\overline{F}$ of each target point over sampled tasks in
semiconductor DQDs and superconducting circuits, respectively. We
can see that as an optimization method with low computational overhead,
even in the case of lack of degrees of freedom, it still performs
well in some certain points. It is worth pointing out that again each
$\overline{F}$ of target point is over 511 preparation tasks, thus
a high-valued $\overline{F}$ implies hundreds of successes of precise
state preparation tasks. While these bad points can also perform better,
in general, by carefully selecting parameters as it is in the case
of single-qubit. For example, if the ``worst point'' in Fig.~\ref{fig:4}
(a) (whose $\overline{F}=0.779$, under $T=10\pi$ and $\mathrm{d}t=\pi/2$)
is performed under $T=18\pi$ and $\mathrm{d}t=\pi/2$, its $\overline{F}$
can also reach 0.813.

\begin{table}
	\centering
\caption{List of parameters and the corresponding average fidelity over all
sampled state preparation tasks in four cases.}
\label{tab1}

\begin{tabular}{|c|c|c|c|c|}

\hline 
Model & Qubit & $T$ & $\mathrm{d}t$ & $\langle\overline{F}\rangle$\tabularnewline
\hline 
\hline 
\multirow{8}{*}{DQDs} & \multirow{4}{*}{Single-} & $\pi$ & $\pi/10$ & 0.951\tabularnewline
\cline{3-5} \cline{4-5} \cline{5-5} 
 &  & 2$\pi$ & $\pi/5$ & 0.973\tabularnewline
\cline{3-5} \cline{4-5} \cline{5-5} 
 &  & 3$\pi$ & $\pi/3$ & 0.977\tabularnewline
\cline{3-5} \cline{4-5} \cline{5-5} 
 &  & 4$\pi$ & $\pi/3$ & 0.983\tabularnewline
\cline{2-5} \cline{3-5} \cline{4-5} \cline{5-5} 
 & \multirow{4}{*}{Two-} & 5$\pi$ & $\pi/2$ & 0.894\tabularnewline
\cline{3-5} \cline{4-5} \cline{5-5} 
 &  & 10$\pi$ & $\pi/2$ & 0.911\tabularnewline
\cline{3-5} \cline{4-5} \cline{5-5} 
 &  & 15$\pi$ & $\pi/2$ & 0.930\tabularnewline
\cline{3-5} \cline{4-5} \cline{5-5} 
 &  & 20$\pi$ & $\pi/2$ & 0.938\tabularnewline
\hline 
\multirow{8}{*}{Superconducting ciruits} & \multirow{4}{*}{Single-} & $\text{\ensuremath{\pi}}$ & $\pi/3$ & 0.983\tabularnewline
\cline{3-5} \cline{4-5} \cline{5-5} 
 &  & $\text{\ensuremath{\pi}}$ & $\pi/5$ & 0.999\tabularnewline
\cline{3-5} \cline{4-5} \cline{5-5} 
 &  & $\text{\ensuremath{\pi}}$ & $\pi/10$ & 0.998\tabularnewline
\cline{3-5} \cline{4-5} \cline{5-5} 
 &  & $\text{\ensuremath{\pi}}$ & $\pi/20$ & 0.999\tabularnewline
\cline{2-5} \cline{3-5} \cline{4-5} \cline{5-5} 
 & \multirow{4}{*}{Two-} & 5$\pi$ & $\pi/4$ & 0.964\tabularnewline
\cline{3-5} \cline{4-5} \cline{5-5} 
 &  & 10$\pi$ & $\pi/4$ & 0.971\tabularnewline
\cline{3-5} \cline{4-5} \cline{5-5} 
 &  & 12$\pi$ & $\pi/4$ & 0.975\tabularnewline
\cline{3-5} \cline{4-5} \cline{5-5} 
 &  & 15$\pi$ & $\pi/4$ & 0.977\tabularnewline
\hline 
\end{tabular}
\end{table}

\subsection{Universal state preparation in noisy environment}

In the previous two subsections, we have explored the performance of our scheme in realizing the universal state preparation neglecting the effects of noises stemming from the surrounding environment and errors arising from systemic imperfections. Now we turn our attention to the robustness of this approach against various adverse factors. 

There are many manifestations of imperfections, and we roughly categorize them into two classes: the static drifts and the dynamic fluctuations. Their impact on the evolution of the system can be taken into account by substituting the pertaining control parameter $f$ with the control-noise term $f+\delta$ or $f+\delta(t)$ in the corresponding Hamiltonian, respectively. 
The static drifts could be brought about by the misaligned control field or constant disturbance from the environment. While the dynamic fluctuations may originate from various time-dependent random control errors and stochastic noises from the environment, such as the no-zero bandwidth of the microwave drive and the charge noises arising from the uncontrolled impurities in the host material \cite{coherent_manipulation_qd,noise1,noise2}. Regardless of their individual statics, these dynamic fluctuations will in concert behave as a noise with normal distribution for the central limit theorem \cite{superconducting_guide}. At each time step the amplitude of the dynamic noise term $\delta(t)$ will be sampled from a zero-mean normal function $N(0,\sigma)$, where $\sigma$ represents the standard deviation and indicates the amplitude of the noise in a sense. For simplicity, the noise term will be kept constant within a time step. We stress here that the noise term is added to the Hamiltonian after the control pulse consequence has been designed to study the performance of our scheme in face of unpredictable noises.
The robustness of our scheme is verified by evaluating the average fidelity $\langle\overline{F}\rangle$ over all of the sampled preparing tasks under different types and strengths of imperfections.

\begin{figure}[htb] 
	\centering
	\subfigure[ ]{\includegraphics[width=7.5cm]{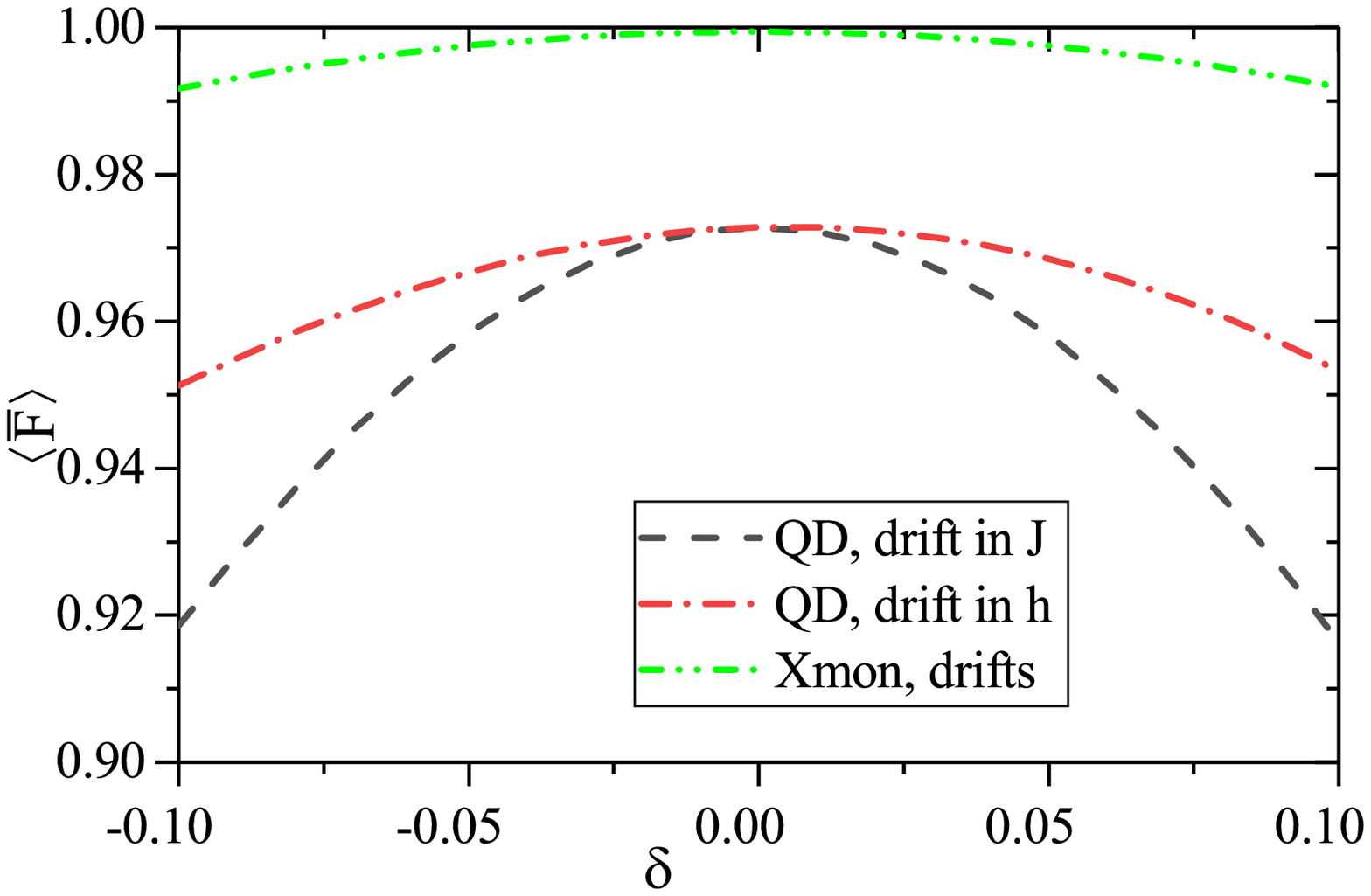}}
	\subfigure[ ]{\includegraphics[width=7.5cm]{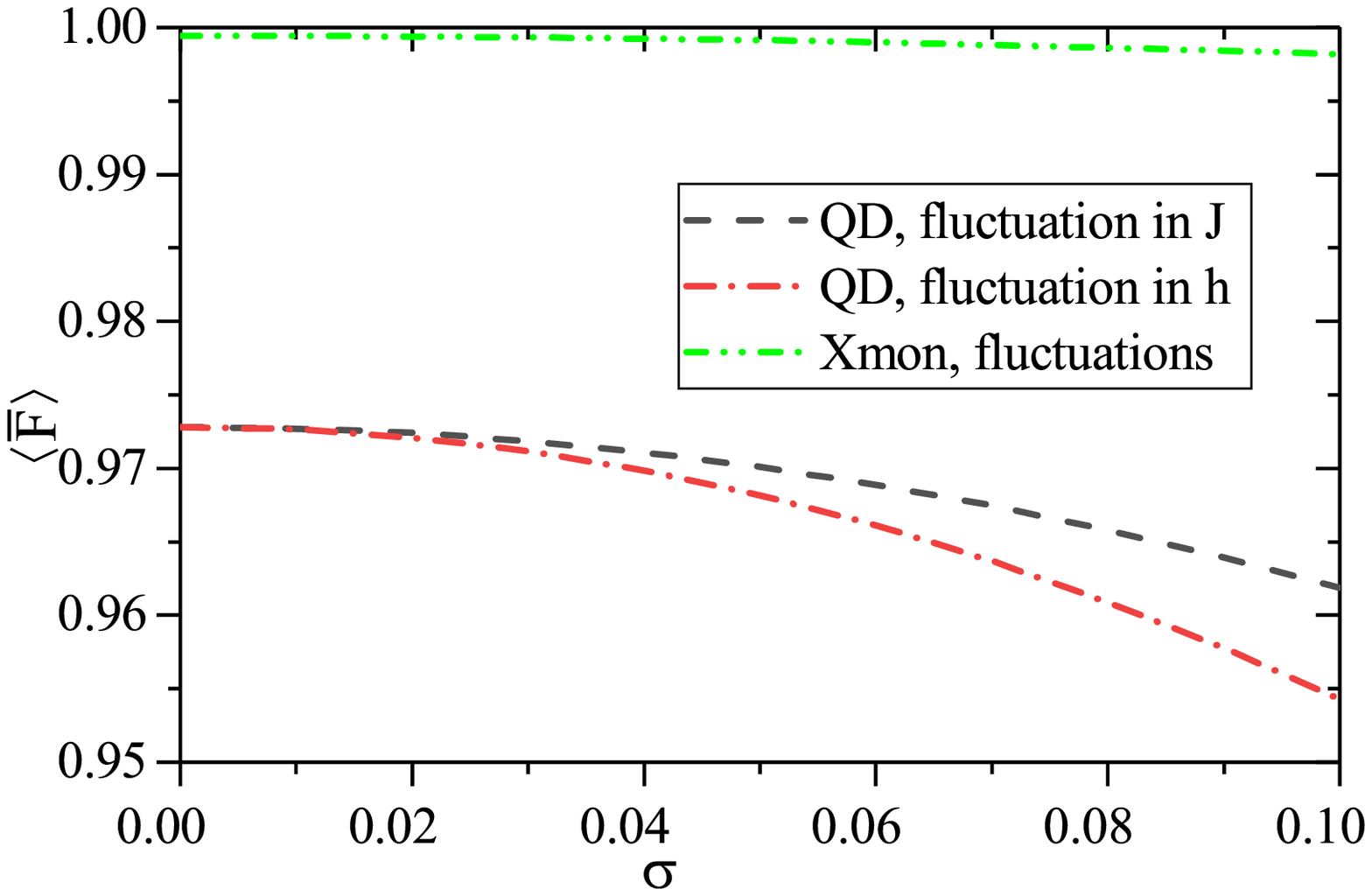}}	
	\subfigure[ ]{\includegraphics[width=7.5cm]{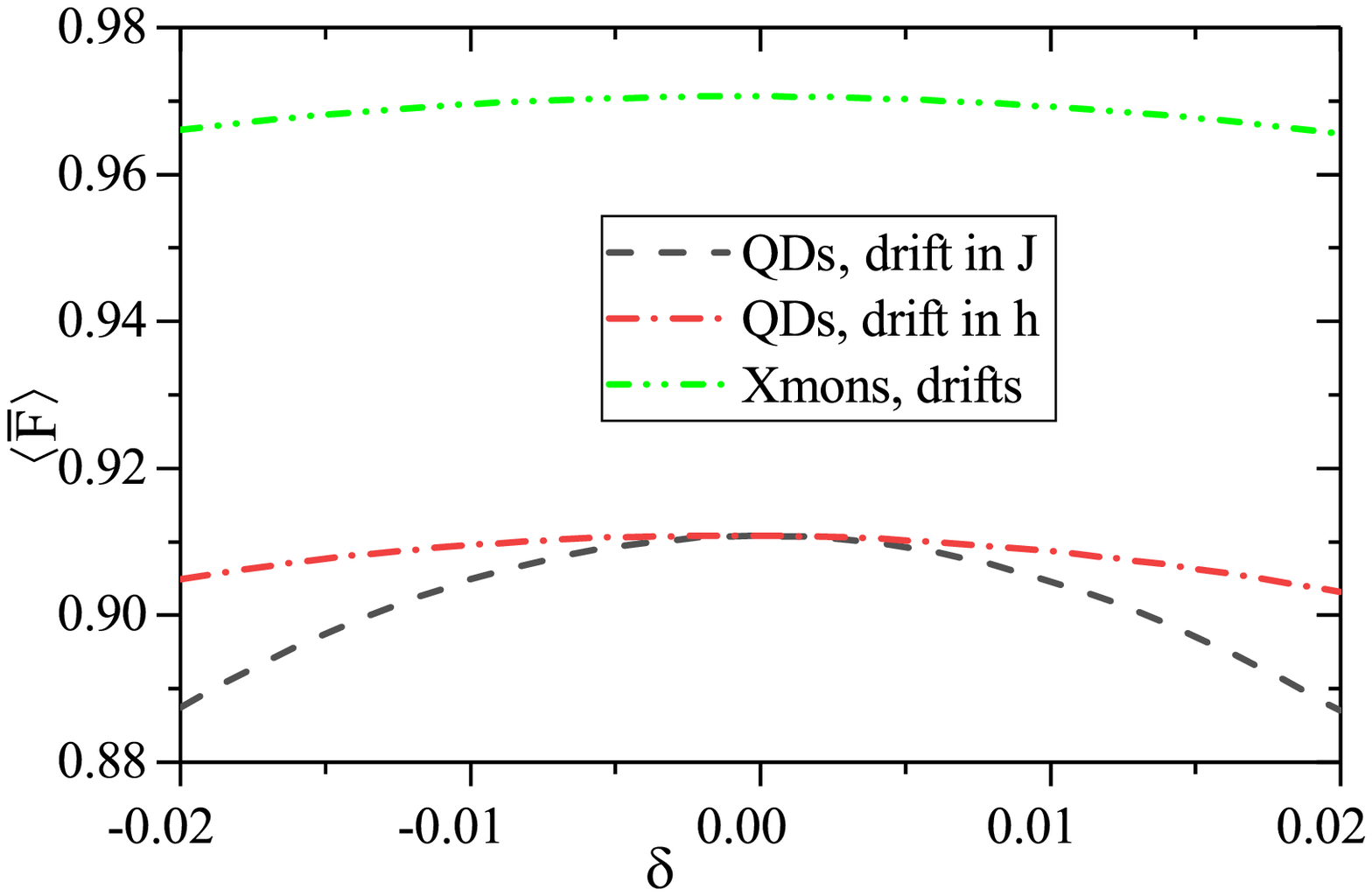}}
	\subfigure[ ]{\includegraphics[width=7.5cm]{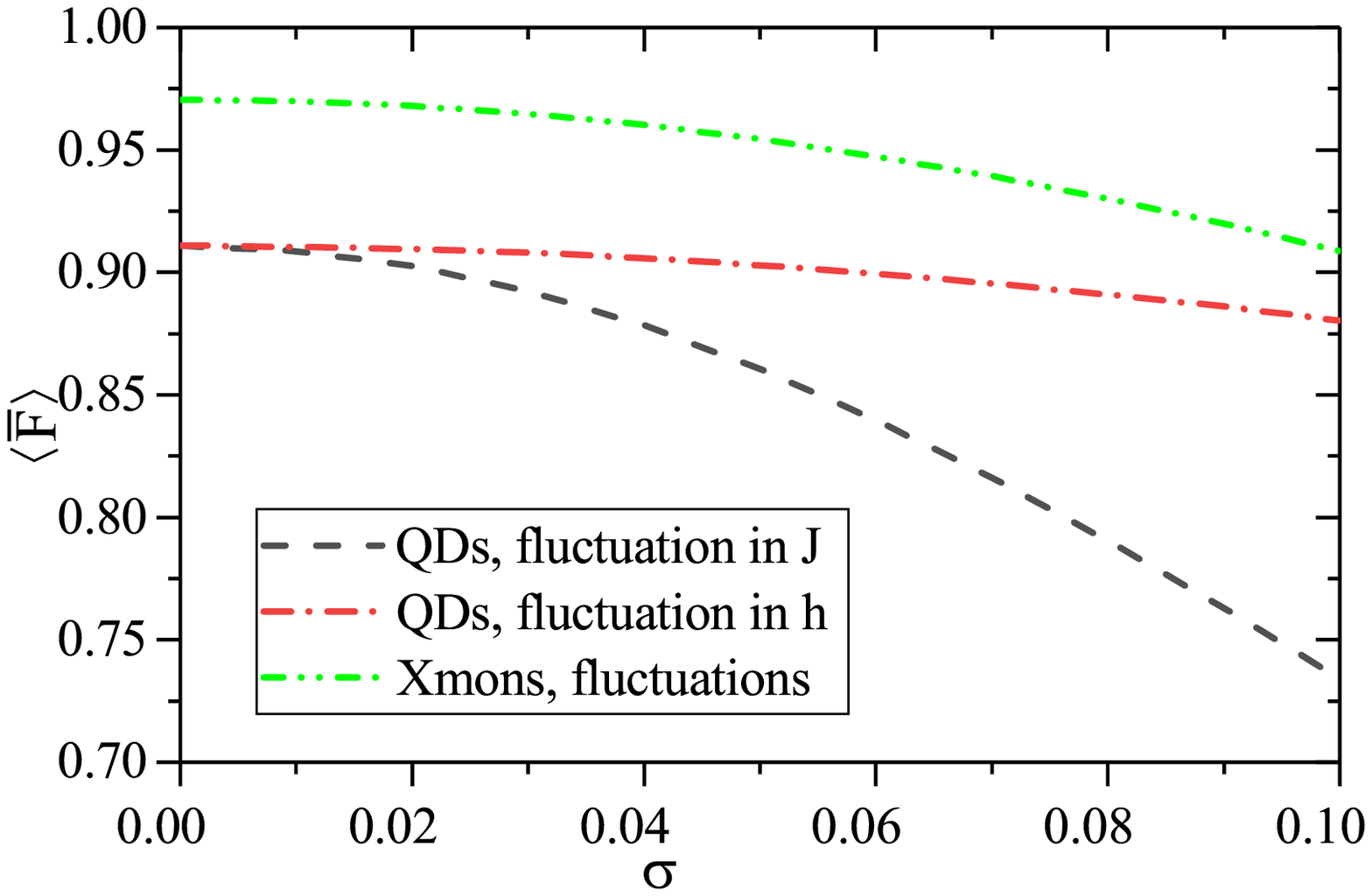}}	
	\caption{Average fidelites of the universal state preparation with RG algorithm over all preparing tasks versus amplitudes of different imperfections: (a) and (b) the static drifts and dynamic fluctuations in the parameters of single-qubit in semiconductor QD system and superconducting Xmon. (c) and (d) the static drifts and dynamic fluctuations in the parameters of two-qubit in semiconductor QDs system and superconducting Xmons.  }
	\label{Fig5}
\end{figure}

For the single-qubit state preparation, the average fidelity $\langle\overline{F}\rangle$ versus the amplitudes of various static drifts and dynamic noises are plotted in Fig.~\ref{Fig5} (a) and (b), respectively.  The impacts of time-dependent errors and noises applied to the QD's exchange coupling $J$ and Zeeman energy gap $h$ are investigated individually. Yet their impacts on Xmon are discussed together for simplicity, i.e., the strengths of imperfections applied to Xmon' control parameters are identified. From Fig.~\ref{Fig5} (a) and (b), we can see that the average fidelity in the QD system is most affected by the static drift in $J$ and the dynamic fluctuation in $h$. While for the superconducting Xmon qubit, overall, the high average fidelity is well maintained even when the system is faced with comparable noises and errors as in the QD system. 

For the two-qubit state preparation, we make the assumption that the static drifts on each qubit are identical i.e., $\delta_{1}=\delta_{2}$ where the subscript refers to the corresponding qubit; while the dynamic noises on two qubits are different, i.e., $\delta_{1}(t)\neq\delta_{2}(t)$ where each $\delta(t)$ is sampled from the same normal function $N(0,\sigma)$ individually.
The average fidelity $\langle\overline{F}\rangle$ of two considered systems as a function of the magnitude of static and dynamic imperfections is showed in Fig.~\ref{Fig5} (c) and (d) respectively. It is obvious that the impact of the static drifts to the systems is similar to the case of single-qubit. However, for the two-qubit QDs system, it is more affected by the dynamic fluctuation in $J$ compared to the single-qubit case.

From the above analysis, we can conclude that, overall, these control trajectories designed by our scheme exhibit a robustness against various errors and noises.

Given the limitations of quantum computing hardware presently accessible,
we simulate quantum computing on a classical computer and generate
the corresponding data. Our algorithms are implemented with PYTHON
3.7.9 and QuTip 4.5.0, and have been run on a 64-core 3.40 GHz CPU with
125.6 GB memory. Details of the running environment of the algorithm
can be found in the Sec.~Data and code availability.

\section{Conclusion  \label{sec:conclusion}}

Precise and universal preparation of single- and two-qubit states 
is fundamental to quantum information and quantum computation. Yet
the difficulty of designing control trajectory in complicated systems
hinders the access to optimal solution for the driving between arbitrary
quantum states. In this work, based on the standard greedy algorithm, we proposed
a revised version, RG, to address this intractable problem. As demonstrations of our scheme, we apply it to the control of single- and two-qubit in the context
of semiconductor DQDs and superconducting circuits and discovered
a well performance, revealing its potential applicability. Compared
with the typical numerical optimizations, our RG algorithm overcomes
the local optimality and achieves a higher preparation
quality. It is also demonstrated that the runtime of designing suited pulses
with our scheme rivals to the GRAPE, which implies an outstanding
efficiency. It outperforms the emerging machine learning approach
with a well accessibility: could tailor the proper control trajectory
to drive arbitrary initial state to other arbitrary target one without
any training but only at a little cost of trial and error. 
We also discover that the control trajectories generated by our scheme are robust against various static and dynamic imperfections.
As a radically different approach from previous methods, our scheme finds
a new route to achieve quantum control optimization.

\section{Models}\label{sec:models}

Among the numerous promising qubit modalities, semiconductor quantum dots \cite{three_two_qubit_quantum_dot,programmable_two_qubit_quantum_dot,Resonantly_driven_quantum_dot,benchmarks_two_qubit_quantum_dot,single_electron_quantum_dot,coherent_manipulation_qd,GaAs_qd,singlet_triplet_qd,quantum_dot_review_guoguoping1,quantum_dot_review_guoguoping2}
and superconducting circuits \cite{superconducting_first,superconducting_quantum_supremacy,Superconducting_An_Outlook,superconducting_a_review,superconducting_guide,superconducting_current_state,qubit_with_josephon_junctions,superconducting_quantum_bits,superconducting_10-qubits,zuchongzhi62,zuchongzhi66}
have captured the imagination of the research field and become the
leading candidates for their desirable merits, e.g. high scalability,
long coherence time and desirable integration with well-established
microfabrication.
In this section, we introduce four models of single- and two-qubit
in semiconductor DQDs and superconducting circuits. 

\subsection{Singlet-triplet qubits in semiconductor double quantum dots}

There are many types of qubit have been proposed and demonstrated
experimentally in semiconductor quantum dots, such as the spin or
charge degrees of the electrons and donor nucleus \cite{Resonantly_driven_quantum_dot,single_electron_quantum_dot,singlet_triplet_qd,quantum_dot_review_guoguoping2,quantum_dot_singlet-triplet_hamiltonian,quantum_dot_review_guoguoping1}.
Due to the merit that it can be driven all electrically, the $S$-$T_{0}$
qubit in DQDs captures the most attention \cite{quantum_dot_singlet-triplet_electrically-control1,quantum_dot_singlet-triplet_electrically-control2,quantum_dot_singlet-triplet_electrically-control3}.
It is encoded by the spins of two electrons trapped in the potential
created by charged electrodes on the surface of the heterostructure
\cite{Resonantly_driven_quantum_dot}.

The effective Hamiltonian of a single $S$-$T_{0}$ qubit driven by
external control field can be written as \cite{coherent_manipulation_qd,singlet_triplet_qd,quantum_dot_singlet-triplet1,quantum_dot_singlet-triplet2,quantum_dot_singlet-triplet_hamiltonian}
\begin{equation}
H(t)=J(t)\sigma_{z}+h\sigma_{x},\label{eq:1}
\end{equation}
under the computational basis $\{|0\rangle,|1\rangle\}$, where $|0\rangle=|S\rangle=(|\uparrow\downarrow\rangle-|\downarrow\uparrow\rangle)/\sqrt{2}$
, $|1\rangle=|T_{0}\rangle=(|\uparrow\downarrow\rangle+|\downarrow\uparrow\rangle)/\sqrt{2}$.
$h$ represents the Zeeman energy gap caused by local inhomogeneous
micromagnetic field. Considering that for the micromagnetic field
resulted by an integrated micromagnet, $h$ is hardly to vary during
runtime, we therefore assume it as a constant and set $h=1$ here
\cite{quantum_dot_singlet-triplet_electrically-control2}. We also
take the reduced Planck constant $\hbar=1$ throughout this work.
The Pauli matrices $\sigma_{x}$, $\sigma_{y}$ and $\sigma_{z}$
indicate rotations about the $x$-, $y$- and $z$-axes of the Bloch
sphere respectively. The exchange coupling $J(t)$ is the only tunable
parameter in this model and can be modulated with electric pulses.
In addition, it is physically restricted to be non-negative and bounded.

The properties of superposition of basis states and entanglement
between multiple qubits are the source of the magic of quantum computing.
In semiconductor DQDs, the interaction Hamiltonian of two entanglement
qubits based on Coulomb interaction reads \cite{quantum_dot_singlet-triplet_electrically-control3,quantum_dot_singlet-triplet_2_qubit,quantum_dot_singlet-triplet_electrically-control1,herunhong_2020_arxiv}
\begin{equation}
\ensuremath{H_{2\!-\!qubit}\!=\!\frac{\hbar}{2}\left(\begin{array}{cccc}
J_{1}\!+\!J_{2} & h_{2} & h_{1} & 0\\
h_{2} & J_{1}\!-\!J_{2} & 0 & h_{1}\\
h_{1} & 0 & J_{2}\!-\!J_{1} & h_{2}\\
0 & h_{1} & h_{2} & -J_{1}\!-\!J_{2}\!+\!2J_{12}
\end{array}\right)},\label{eq:2}
\end{equation}
under the basis constituted by $\ensuremath{\{|SS\rangle,|ST_{0}\rangle,|T_{0}S\rangle,|T_{0}T_{0}\rangle\}}.$
$h_{i}$ and $J_{i}$ are the Zeeman energy gap and exchange interaction
with the subscript $i=1,2$ referring to the corresponding qubit.
The coupling strength between two qubits $\ensuremath{J_{12}\propto J_{1}J_{2}}$.
To maintain the entanglement between two qubits, it has to keep $J_{i}>0$.
We set $h_{1}=h_{2}=1$ and $J_{12}=J_{1}J_{2}/2$ here for simplicity.

\subsection{Superconducting quantum circuits qubits}

As a kind of ``artificial atom'', the Hamiltonian of superconducting
qubits can be designed just by tailoring the capacitance, inductance
and Josephson energy \cite{superconducting_a_review}. According
to the degrees of freedom and the ratio of Josephson energy to charging
energy, superconducting qubits can mainly be classified into three
categories: charge qubits \cite{superconducting_first}, flux qubits
\cite{superconducting_flux_qubit} and phase qubits \cite{superconducting_phase_qubit1,superconducting_phase_qubit2}.
Based on the above three archetypes, a variety of new types of superconducting
qubits emerges, such as Transmon \cite{transmon_qubit}, Xmon
\cite{x-mon_paper,x-mon_1,x-mon_2}, Gmon \cite{g_mon} and
so on. Considering the representativeness of the Xmon type superconducting
qubits, we take it as an example here. Nonetheless, this control scheme
is also applicable to other superconducting qubit models.

When the qubit resonates with the microwave drive, the Hamiltonian
of a single Xmon qubit in the rotating frame can be written as
\cite{superconducting_guide,x-mon_paper,x-mon_1,superconducting_quantum_computing_a_review}
\begin{equation}
\ensuremath{H=\frac{\hbar}{2}A(\mathrm{cos}\phi\sigma_{x}+\mathrm{sin}\phi\sigma_{y})},
\end{equation}
where $A$ and $\phi$ are the amplitude and phase of the microwave,
respectively. Obviously, when $\phi=0$, $H=\frac{\hbar}{2}A_{x}\sigma_{x}$;
in contrast, when $\phi=\pi/2$, $H=\frac{\hbar}{2}A_{y}\sigma_{y}$.
Thus, the rotations about the $x$- and $y$-axes of the Bloch sphere
can be obtained by properly setting $A$, $\phi$ and the duration
$\tau$ of the microwave. In addition, the operation about the $z$-axis can be implemented physically by adjusting the current flowed
into the superconducting quantum interference device loop through the so-called $Z$-line and the Hamiltonian
can be expressed as \cite{superconducting_guide,superconducting_quantum_computing_a_review}
\begin{equation}
\ensuremath{H=-\frac{\hbar}{2}A_{z}\sigma_{z},}
\end{equation}
when the $XY$ microwave drive is absent. $A_{z}$ is determined by
the structure of the qubit as well as the current intensity and is
bounded to be nonnegative and finite.

For two capacitively coupled Xmon qubits with same frequency,
using rotating-wave approximation, the additional interaction
term can be written as \cite{superconducting_a_review,superconducting_guide,superconducting_quantum_computing_a_review}
\begin{equation}
H_{couple}=\hbar g(\sigma_{1}^{+}\sigma_{2}^{-}+\sigma_{1}^{-}\sigma_{2}^{+}),
\end{equation}
where $\sigma_{j}^{\pm}=\frac{1}{2}(\sigma_{x}^{j}\pm i\sigma_{y}^{j})$
with the superscript $j\in\{1,2\}$ referring to the corresponding qubit.
$g$ is the coupling strength and we set $g=1$ here.

\section*{Data and code availability \label{sec:data-and-code}}
The code, running environment of algorithm and all data used or presented
in this paper are available from the corresponding author upon reasonable
request or from Gitee in (\url{https://gitee.com/herunhong/USP-via-RG}).

\ack
The first two authors Run-Hong He and Hai-Da Liu have contributed equally to this work. 
The author Run-hong He would also like to personally thank Jing-Hao Sun and Chen-Chen for
useful discussions. This work was supported by the Natural Science Foundation
of China (Grant Nos. 11475160, 61575180), and the Natural Science Foundation
of Shandong Province (Grant No. ZR2014AM023). 

\section*{References}

\bibliographystyle{unsrt}
\bibliography{References}
\end{document}